# Efficient Service Broker Policy For Large-Scale Cloud Environments

**Mohammed Radi**

**Computer Science Department, Faculty of Applied Science**
**Alaqsa University, Gaza Palestine**
**Moh_radi@alaqsa.edu.ps**

*Abstract*

Algorithms, policies, and methodologies are necessary to achieve high user satisfaction and practical utilization in cloud computing by ensuring the efficient and fair allocation of every computing resource. Whenever a new job arrives in cloud environments, the service broker is responsible for selecting the data center that will execute that job. Selecting data centers serves an important function in enhancing the performance of a cloud environment. This study proposes a new service broker policy for large-scale cloud applications based on the round-robin algorithm. The proposed policy is implemented and evaluated using a CloudAnalyst simulator. It is then compared with three existing policies in terms of overall average response time by using different virtual machine load balancing algorithms. Simulation results show that the proposed policy improves the overall average response time relative to that of the other policies.

## 1. Introduction

Cloud computing is a large-scale distributed computing paradigm that is driven by economies of scale, in which a pool of abstracted, virtualized, and dynamically scalable managed computing power, storage, platforms, and services are delivered on demand to external customers over the Internet [1, 2]. Progress is being made in cloud computing every day.

Cloud computing is designed to provide service rather than a product. Transparency is one of the main design issues of cloud computing, in which services such as computation, software, data access, and storage are provided to users without the knowledge of the physical location and configuration of the server that provides such services [3].

Given the progress of cloud services, large-scale software systems, such as social networking sites and e-commerce applications, are gaining popularity. These software systems greatly utilize cloud services to minimize costs and improve service quality to end users. Several factors affect the cost and quality of the cloud. Some of these factors are the distribution (geographic) of the user bases, the availability of the Internet infrastructure within those geographic areas, the dynamic nature of the usage patterns of the user base, and how well cloud services can be adapted or dynamically reconfigured.

Some studies have used simulation techniques to investigate the behavior of large-scale distributed systems [4, 5]. These studies have shown that cloud analysis [6] can be used for simulating this type of large-scale applications.

Cloud computing has different metrics, including fault tolerance, availability, scalability, flexibility, reduced overhead for users, performance, and on-demand services. Algorithms, policies, and methodologies are necessary to achieve high user satisfaction and resource utilization in cloud computing by ensuring an efficient and fair allocation of every computing resource. Selecting proper service broker policies and scheduling algorithms in the cloud environments of large-scale applications serves an essential function in the system performance [4, 5]. Thus, a technique must be designed to minimize the response time, cost, resource utilization, and overhead by distributing the user application workload among various data centers [7].

Driven by the preceding considerations, this study proposes and implements a new service broker policy (data center selection) based on the round-robin (RR) algorithm to minimize the service response time. The proposed policy is evaluated and compared with three existing policies using the CloudAnalyst simulator.

The rest of the paper is organized as follows: Section 2 discusses a number of previous works related to cloud service broker policies. Section 3 describes the Cloud Analyst tool. Section 4 describes the virtual machine (VM) load balancing policy. Section 5 discusses the proposed policy. Section 6 presents the experimental results and discussion. A brief conclusion ends the study.

## 2. Related Works

In cloud computing, the geographical location of the target data center has a significant influence on overall performance of cloud system because of different network issues and the location of user groups [8]. The service broker policy is responsible for routing the requests of users originating from different user groups, which are located in different geographical regions in the globe, to the data centers in cloud. The data centers are also distributed in different geographical regions.

The standard service broker policy provided by Cloud Analyst can be any of the following types:

- **Service Proximity-Based Broker Policy**

This policy is considered the shortest path data center. The service broker sends the request to the closest data center in consideration of network latency.

- **Performance Optimization Policy**

In this policy, the service broker actively monitors all data centers and sends the request to the data center that provides the best response time to the end user at the time of query.

- **Dynamic Configuration Policy**

In this policy, the service broker is assigned the additional job of scaling the application deployment depending on the load it is currently facing. The service broker dynamically increases or decreases VMs in the data centers according to the current processing times relative to the best processing time that has ever been achieved.

A number of studies have been conducted on the effect of service broker policy on the performance of cloud environments [4, 5, 8]. [4, 5] used CloudAnalyst to represent the performance analysis of three previous broker policies in combination with different load balancing policies for large-scale applications using different infrastructural environments. They concluded that the service broker policy affects the overall response time of the system. Although P. M. Rekha et. at. [8] analyzed various service broker policies for selecting data centers and compared their cost, the idea of service brokering in a cloud continues to be the subject of many research works.

A. Aikar et al. [9] proposed an effective data center selection algorithm for a federated cloud, in which the data center is selected based on the matrix values. The matrix is assigned to each region. This matrix contains information on the cost required for the resources of a data center and distance of the request location.

A number of research works extended the service proximity-based broker policy [3, 10, 11] and studied the case in which more than one data center is located in the selected region. D. Kapgate [12] mainly focused on implementing the predictive service broker algorithm based on the weighted moving average forecast model. This research shows how the proposed predictive service broker algorithm minimizes the response time experienced by users and the load on the data center.

## 3. Introduction to CloudAnalyst

CloudSim enables smooth modeling, simulation, and experimenting on cloud computing infrastructure [13]. It is a platform that can be used to model data centers, service brokers, and scheduling and allocation policies of large-scale cloud platforms. CloudAnalyst [6] is built directly on top of the CloudSim toolkit.

The following are the main features of CloudAnalyst:

- easy-to-use graphical user interface (GUI)
- ability to define a simulation with a high degree of configurability and flexibility
- repeatability of experiments
- graphical output
- use of consolidated technology and ease of extension

The following are the main components of CloudAnalyst and the function of each component:
- **GUI Package.** It is responsible for the GUI, for serving as the front end controller for the application, and for managing screen transitions and other user interface activities.
- **Region.** Six regions correspond to six continents in the world.
- **User Base.** This component models a group of users and generates traffic that represents the users.
- **Data Center.** It encapsulates a set of computing hosts or servers that are either heterogeneous or homogeneous in nature depending on their hardware configurations.
- **Data Center Controller.** This component controls data center activities.
- **Cloudlet.** It specifies a set of user requests. It contains the application ID, name of the user base as originator for routing back the responses, size of request execution commands, and input and output files.
- **Internet.** This component models the Internet and implements the traffic routing behavior.
- **Internet Characteristics.** This component is used to define the characteristics of the Internet applied during simulation.
- **VM Load Balancer.** This component models the load balance policy used by data centers when the serving allocation is requested.

- **Service Broker.** The service broker determines which data center should be selected to provide services to requests from the user base.

## 4. VM Load Balancing Policy

The VM allocation policy is used to route user requests received in the form of cloudlet to the VM for processing. The main three algorithms considered in this study are defined as follows[14]:

### 4.1 RR Policy
RR is a straightforward policy in which the requests of the clients are handled in a circular manner basis of a first-come first-served .

### 4.2 Throttled Policy
In this policy, each VM processes only one job at a time; a new job can be processed only when the current job is completed successfully. The load balancer entity maintains an index table of all the VMs and their current states (available or busy). As soon as the data center controllers query the load balancer for the allocation of the VM, the load balancer searches the index table for an available VM. If the load balancer finds an available VM, the load balancer returns the VM ID to the data center controller. Otherwise, the load balancer simply returns null. If the data center controller receives null from the load balancer, the request is queued until a VM becomes available.

### 4.3 Active Monitoring (AM) Policy
In this policy, the load balancer actively monitors the load on all the VMs and distributes the load equally among them. The load balancer maintains an index table of the VMs and the number of allocations assigned to each VM. When a new job arrives, the data center controller queries the load balancer for the allocation of a new VM. The load balancer searches the index table for the VM with the least load. When the load balancer finds the VM with the least load, it returns the VM ID to the data center controller.

## 5. Service Broker Policy

The service broker is responsible for selecting the data center that will handle the cloudlet, while the VM load balancer is responsible for the load balance in the cloudlet among the VMs. The process of routing the user request, including the use of the service broker policy and VM load balancer, is shown in Figure 1 [6] [13].

The user base generates an Internet cloudlet. The Internet refers to the service broker for data center controller selection. The service broker uses the service broker policy to return the information on the selected data center controller to the Internet. Using this information, the Internet sends the REQUEST to the data center controller. The selected data center controller uses the VM load balancer, and the requests are processed. The selected data center controller then sends the RESPONSE to the Internet.

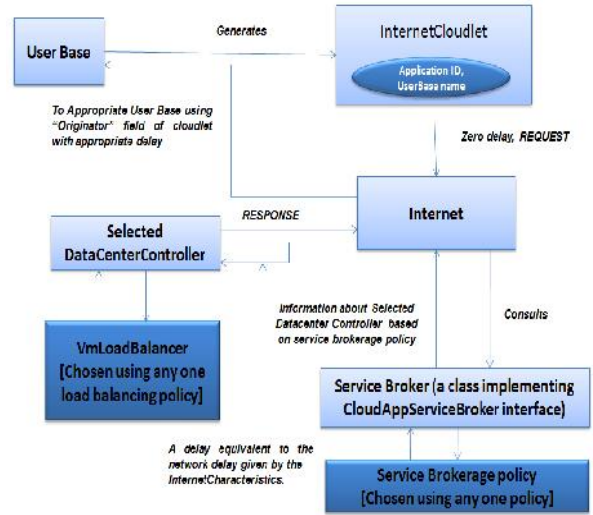

Figure 1: Routing of user request

## 5.1 Proposed Service Broker Policy

The proposed service broker policy is based on the traditional RR algorithm. In the RR policy, the requests of the clients are handled in a circular manner basis of a first-come first-served. The Internet directly interacts with the requests of the users, and the service broker is the entity that assigns load to the data center. In this policy, the service broker keeps a record of all the available data centers and keeps track of the next data center to which the next job is supposed to be assigned. When a new request from a user arrives, the Internet forwards the request to the service broker. The service broker selects the next data center in a circular order and assigns the job in circular manner. The details of the algorithms are shown in Figure 2.

---

**Service Broker Based on RR Algorithm**

**Input:** All Data Centers
**Output:** Target Data Center

CurrentDataCenter_id ← CurrentDataCenter_id +1
 **if** CurrentDataCenter_id > = allDataCenters.size() **then**
　　CurrentDataCenter_id ← 0;
 **end if**
**Return** Current_data_center_id

---

Figure 2: RR service broker policy

For example, if only one data center exists, then all the cloudlets are directly assigned to it. In this case, two data centers exist (DC1 and DC2). The first cloudlet is assigned to DC1, and DC2 is assigned as the target data center for the second cloudlet, the third cloudlet is assigned to DC1, and so on. If three data centers exist, then the first cloudlet is sent to DC1, the second cloudlet to DC2, the third cloudlet to DC3, the fourth cloudlet to DC1, the fifth cloudlet to DC2, and so on.

## 6. Experimental Results and Discussions

Large-scale applications that benefit from cloud computing include social networking applications, e-commerce applications, and online education applications. The present study considers the social networking application Facebook. In our experiments, we define six user bases representing six regions, with the parameters presented in Table 1. The other simulation parameters that are used are presented in Table 2.

Table 1. Region Parameters fixed for simulations

| Users During Off peak Hours | Users During Peak Hours | Peak Hour (GMT) | Region | User Base |
|---|---|---|---|---|
| 1104866 | 11048660 | 13:00–15:00 | R0 | UB1 |
| 562655 | 5626555 | 15:00–17:00 | R1 | UB2 |
| 1164178 | 11641787 | 20:00–22:00 | R2 | UB3 |
| 1076411 | 10764114 | 01:00–03:00 | R3 | UB4 |
| 201027 | 2010279 | 21:00–23:00 | R4 | UB5 |
| 67986 | 679869 | 09:00–11:00 | R5 | UB6 |

Table 2. Parameters fixed for simulations

| Value | Parameter |
|---|---|
| 60 min | Simulation duration |
| 12 | Requests per user per hour |
| 100 bytes | Data size per request per hour |
| 40 | Number of hosts (each with four processors) |
| 10000 | User grouping factor in user bases |
| 1000 | Request grouping factor in data centers |
| 500 bytes | Executable instruction length per request |

In the experiments, the effects of the proposed RR policies were investigated and compared with those of the other policies. The overall average response time was the main performance metrics. The simulation runs were conducted for 72 times. Three variants were changed each time: the number and distribution of data centers, the service broker policy, and the VM load balancing policy. The number and distribution of the data centers were changed according to the following combinations: (R0 and R2), (R1 and R2), (R1 and R3), (R0, R1, and R2), (R0, R2, and R4), and (R1, R3, and R5). The service broker policies were RR, closest data center, optimal response time, and dynamic configuration. The VM load balancing policies were RR, AM, and throttling. The observations on the overall average response time during all the simulation runs are depicted in Figure 3, Figure 4, and Figure 5.

Figure 3 compares the overall average response time of the four service broker polices using RR as the load balancing policy at the VM level. Figure 4 compares the overall average response time of the four service broker polices using AM. Figure 5 compares the overall average response time of the four service broker polices using the throttling policy.

In Figure 3, Figure 4, and Figure 5, the horizontal axis indicates the data center distributions, and the vertical axis the overall average response time (in msec).

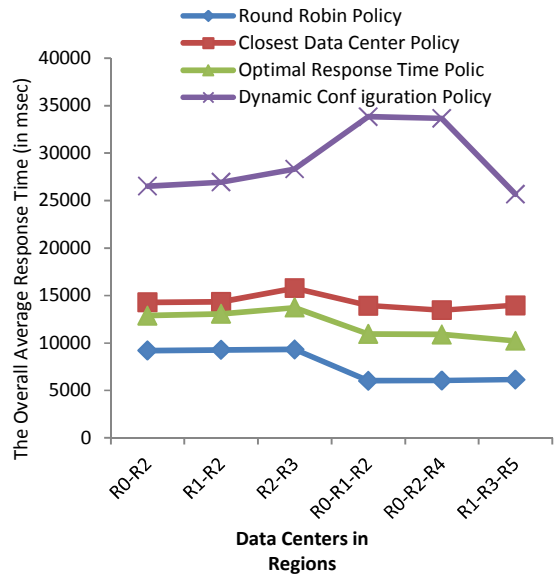

Figure 3: Overall response time of different number of data centers located at different regions using different policies at the broker level and RR load balancing policy at the VM level.

The results in Figure 3, Figure 4, and Figure 5 show a drastic reduction in the average overall response time observed by the user in the proposed policy relative to that in the traditional service broker policies, while using a different load balancing policy at the VM level.

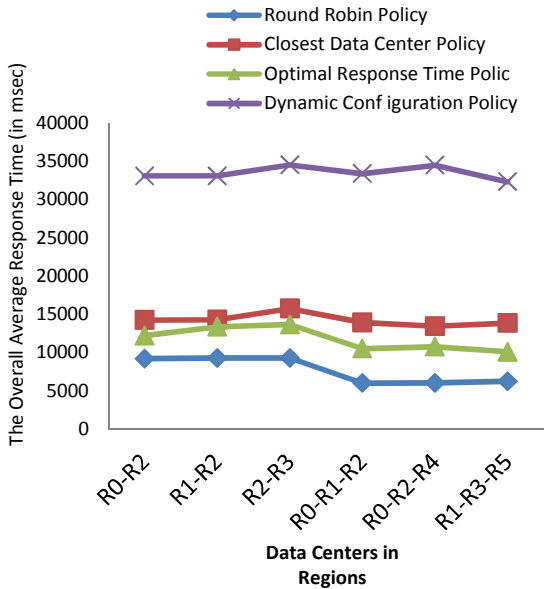

Figure 4: Overall response time of different number of data centers located at different regions using different policies at the broker level and AM load balancing policy at the VM level.

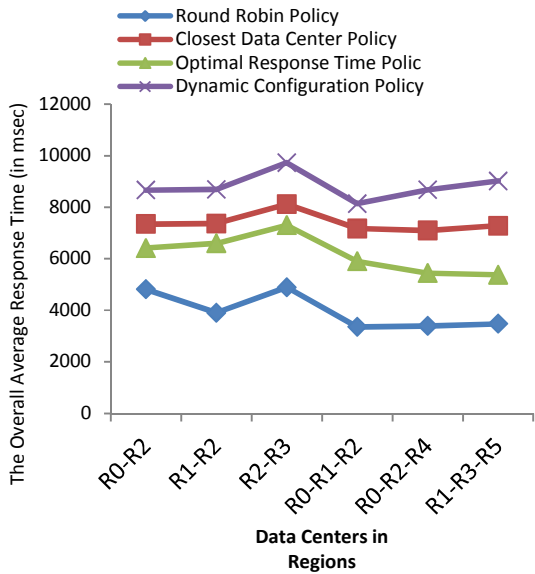

Figure 5: Overall response time of different number of data centers located at different regions using different policies at the broker level and throttled load balancing policy at the VM level.

The proposed policy guarantees a fair distribution of the user requests to all data centers. This fair distribution significantly improves the response using the proposed policy. Unlike the traditional service broker policies (closest data center policy, optimal response time, and dynamic configuration policy), the proposed policy exhibits a drastic reduction in the average overall response time while using different load balancing policy at the VM level, as discussed in the section "Experimental Results and Discussions." We recommend using the proposed service broker policy in large-scale cloud environments.

## 7. Conclusions

Given the features of a large-scale application cloud computing, a service broker (data center selection) policy based on the RR algorithm is proposed. The proposed policy is evaluated and compared with existing policies using the CloudAnalyst simulator. The simulation experiment results show a drastic improvement in the average overall response time. In the future, we will investigate the proposed policy under different configurations and with different VM load balancers.

**Mohammed Radi** has completed his PhD degree at 2009 in distributed database from University Putra Malaysia. His MSc was at 2003 in computer science from University of Jordan and BSc in 2001 in computer science from AlAzhar University Gaza. Currently he is an Assistant professor ant Al-Aqsa University - Gaza. His primary research interest is could compute, distributed system and distributed database.